\documentclass[sigconf]{acmart}

\usepackage{booktabs} 
\usepackage{url}
\usepackage{graphicx}
\usepackage{subcaption}
\usepackage[show]{notes-alt}
\usepackage{balance}
\usepackage{enumitem}
\usepackage{booktabs} 
\usepackage{balance}

\newcommand\cnt{\textsc{Citation Need}}
\newcommand\crt{\textsc{Citation Reason}}
\newcommand\crtax{Citation Reason Taxonomy}
\copyrightyear{2019}
\acmYear{2019} 
\setcopyright{iw3c2w3}
\acmConference[WWW '19]{Proceedings of the 2019 World Wide Web Conference}{May 13--17, 2019}{San Francisco, CA, USA}
\acmBooktitle{Proceedings of the 2019 World Wide Web Conference (WWW '19), May 13--17, 2019, San Francisco, CA, USA}
\acmPrice{}
\acmDOI{10.1145/3308558.3313618}
\acmISBN{978-1-4503-6674-8/19/05}

\begin{document}
\title{Citation Needed: A Taxonomy and Algorithmic Assessment of Wikipedia's Verifiability}

\author{Miriam Redi}
\orcid{0000-0002-0581-0251}
\affiliation{%
  \institution{Wikimedia Foundation}
  \institution{London, UK}
  }

\author{Besnik Fetahu}
\affiliation{%
  \institution{L3S Research Center\\Leibniz University of Hannover}
}

\author{Jonathan Morgan}
\affiliation{%
  \institution{Wikimedia Foundation}
  \institution{Seattle, WA}
  }

\author{Dario Taraborelli}
\affiliation{%
  \institution{Wikimedia Foundation}
  \institution{San Francisco, CA}
  }

\begin{abstract}

Wikipedia is playing an increasingly central role on the web, and the policies its contributors follow 
when sourcing and fact-checking content affect 
million of readers. Among these core guiding principles, \emph{verifiability} policies have a particularly important role. Verifiability requires that information included in a Wikipedia article be corroborated against \emph{reliable secondary sources}. Because of the manual labor needed to curate and fact-check Wikipedia at scale, however, its contents do not always evenly comply with these policies. Citations (i.e. reference to external sources) may not conform to 
verifiability requirements or may be missing altogether, potentially weakening the reliability of specific topic areas of the free encyclopedia.
In this paper, we aim to provide an empirical characterization of the reasons \emph{why} and \emph{how} Wikipedia cites external sources to comply with its own verifiability guidelines. First, we construct a \emph{taxonomy of reasons} why inline citations are required by collecting labeled data from editors of multiple Wikipedia language editions. We then 
collect a large-scale crowdsourced dataset of Wikipedia sentences annotated with categories derived from this taxonomy. Finally, we design and evaluate algorithmic models to determine if a statement \emph{requires a citation}, and to predict the \emph{citation reason} based on our taxonomy. We evaluate the robustness of such models across different classes of Wikipedia articles of varying quality, as well as on an additional dataset of claims annotated for fact-checking purposes. 
\end{abstract}

%
%
%
%

\ccsdesc[300]{Computing methodologies~Neural networks}
\ccsdesc[300]{Computing methodologies~Natural language processing}
\ccsdesc[300]{Information systems~Crowdsourcing}
\ccsdesc[300]{Human-centered computing~Wikis}

\keywords{Citations; Data provenance; Wikipedia; Crowdsourcing; Deep Neural Networks;}

\maketitle

\section{Introduction}
Wikipedia is playing an increasingly important role as a ``neutral'' arbiter of the factual accuracy of information published
in the web.
Search engines like Google 
systematically pull content from Wikipedia 
and display it alongside search results~\cite{mcmahon2017substantial}, while large social 
platforms 
have started experimenting with links to Wikipedia articles, in an effort to tackle the spread of disinformation 
~\cite{matsakis2018}. 

Research on the accuracy 
of information available on Wikipedia suggests that despite its radical openness---anyone can edit most articles, often without having 
an account 
---the confidence that other platforms place in the factual accuracy of Wikipedia 
is largely justified. Multiple studies have shown that Wikipedia's content across 
topics is of a generally high quality\cite{lavsa2011reliability,giles2005internet}, that the vast majority of vandalism 
contributions are quickly corrected~\cite{Priedhorsky:2007:CDR:1316624.1316663,Geiger:2013:LBW:2491055.2491061,Kumar:2016:DWI:2872427.2883085}, and that Wikipedia's decentralized
process for vetting information works effectively even under conditions where reliable information is hard to come by,
such as in 
breaking news events~\cite{keegan2013news}.

Wikipedia's editor communities govern themselves 
through a set of collaboratively-created policies and guidelines~\cite{beschastnikh2008wikipedian,forte2009consensus}. %
Among those, the Verifiability policy\footnote{\url{https://en.wikipedia.org/wiki/Wikipedia:Verifiability}} is a key mechanism that allows Wikipedia to maintain its quality. Verifiability mandates that, in principle, ``all material in Wikipedia... articles must be verifiable'' and attributed to reliable secondary sources, ideally through inline citations, and that unsourced material should be removed or challenged with a \textit{\{citation needed\}} flag.

While the role citations serve to meet this requirement is straightforward, the process by which editors determine which claims require citations, and why those claims need citations, are less well understood. In reality, almost all Wikipedia articles contain at least some unverified claims, and 
while high quality articles may cite hundreds of sources, 
recent estimates suggest that the proportion of articles with 
few or no references can be substantial~\cite{lewoniewski2017}. While as of February 2019 there exists more than $350,000$ articles with one or more \textit{\{citation needed\}} flag, we might be missing many more.

Furthermore, previous research suggests that editor citation practices are not systematic, but often contextual and ad hoc. Forte et al. ~\cite{Forte:2018:IFO:3148330.3148347} demonstrated that Wikipedia editors add citations primarily for the purposes of ``information fortification'': adding citations to protect information that they believe may be removed by other editors. Chen et al.~\cite{Chen:2012:CND:2462932.2462943} found evidence that editors often add citations to existing statements relatively late in an article's lifecycle. We submit that by understanding the reasons why editors prioritize adding citations to some statements over others we can support the development of systems to scale volunteer-driven verification and fact-checking, potentially increasing Wikipedia's long-term reliability and making it more robust against information quality degradation and coordinated disinformation campaigns. 
 

Through a combination of qualitative and quantitative methods, we conduct a systematic assessment of the application of Wikipedia's verifiability policies at scale. We explore this problem throughout this paper by focusing on two tasks: 
\begin{enumerate}[leftmargin=*]
\item\cnt: identifying \textit{which} statements need a citation.
\item\crt: identifying \textit{why} a citation is needed. 
\end{enumerate}
By characterizing qualitatively and algorithmically these two tasks, this paper makes the following contributions:
\begin{itemize}[leftmargin=*]
    \item We develop a \crtax{}\footnote{We use here the term "taxonomy" in this context as a synonym of \textit{ coding scheme}.} describing reasons why individual sentences in Wikipedia articles require citations, based on verifiability policies as well as labels collected from editors of the English, French, and Italian Wikipedia (See Sec. \ref{sec:taxonomy}). 
    \item We assess the validity of this taxonomy and the corresponding labels through a crowdsourcing experiment, as shown in Sec. \ref{sec:dataset}. We find that sentences needing citations in Wikipedia are more likely to be  historical facts, statistics or direct/reported speech. We publicly release this data as a Citation Reason corpus.
    \item We train a deep learning model to perform the two tasks, as shown in Secc. \ref{sec:citation_needed_approach} and \ref{sec:citation_reason}. We demonstrate the high accuracy (F1=$0.9$) and generalizability of the \textsc{Citation Need} model, explaining its predictions by inspecting the network's attention weights. 
\end{itemize}
These contributions open a number of further directions, both theoretical and practical, that go beyond Wikipedia and that we discuss in Section \ref{sec:discussion}.

\section{Related Work}



The contributions described in this paper build on three distinct bodies of work: crowdsourcing studies comparing the judgments of domain experts and non-experts, machine-assisted citation recommendations on Wikipedia, and automated detection and verification of factual claims in political debates.

\vspace{2pt}
\noindent\textbf{Crowdsourcing Judgments from Non-Experts.} 
Training machine learning models to perform the \textsc{citation need} and \textsc{citation reason} tasks requires large-scale data annotations.
While generating data for the first task necessarily requires expert knowledge (based on understanding of policies), we posit that defining the \emph{reasons} why a citation that has already been deemed appropriate is needed can be effectively performed by people without domain expertise, such as crowdworkers. 

Obtaining consistent and accurate judgments from untrained crowdworkers can be a challenge, particularly for tasks that require contextual information or domain knowledge. 
However, a study led by Kittur~\cite{Kittur:2008:CUS:1357054.1357127} found that crowdworkers were able to provide article quality assessments that mirrored assessments made by Wikipedians by providing clear definitions and instructions, and by focusing the crowdworkers attention on the aspects of the article that provided relevant evaluation criteria. Similarly, Sen et al. ~\cite{Sen:2015:TSA:2675133.2675285} demonstrated that crowdworkers are able to provide semantic relatedness judgments as scholars when presented with keywords related to general knowledge categories. 

Our labeling approach aims to assess whether crowdworkers and experts (Wikipedians) agree in their understanding of verifiability policies---specifically, whether non-experts can provide reliable judgments on the reasons why individual statements need citations.


\vspace{2pt}
\noindent\textbf{Recommending Sources.} 
Our work is related to a body of bibliometrics works on citation analysis in academic texts. These include unsupervised methods for citation recommendation in articles \cite{he2011citation}, and supervised models to identify the purpose of citations in academic manuscripts\cite{abu2013purpose}. Our work explores similar problems in the different domain of Wikipedia articles: while scholarly literature cites work for different purposes\cite{abu2013purpose} to support original research, the aim of Wikipedia's citations is to verify existing knowledge.

Previous work on the task of source recommendation in Wikipedia has focused on cases where statements are marked with a \emph{citation needed} tag~\cite{DBLP:conf/acl/SauperB09,DBLP:conf/cikm/FetahuMA15,DBLP:conf/cikm/FetahuMNA16,DBLP:conf/emnlp/FetahuMA17}. Sauper et al.  \cite{DBLP:conf/cikm/FetahuMA15,DBLP:conf/acl/SauperB09} focused on adding missing information in Wikipedia articles from external sources like news, where the corresponding Wikipedia entity is a salient concept. In another study \cite{DBLP:conf/cikm/FetahuMNA16}, Fetahu et al. used existing statements that have either an outdated citation or \emph{citation needed} tag to query for relevant citations in a news corpus. Finally, the authors in \cite{DBLP:conf/emnlp/FetahuMA17}, attempted to determine the \emph{citation span}---that is, which parts of the paragraph are covered by the citation---for any given existing citation in a Wikipedia article and the corresponding paragraph in which it is cited. 

None of these studies provides methods to determine whether a given (untagged) statement \emph{should} have a citation and \textit{why} based on the citation guidelines of Wikipedia. 

\begin{table*}[t]
\resizebox{0.92\textwidth}{!}{
\begin{tabular}{ll}
\multicolumn{2}{l}{\textbf{Reasons why citations are needed}}  \\ 
\midrule
\textit{Quotation}   & The statement appears to be a direct quotation or close paraphrase of a source    \\
\textit{Statistics}  & The statement contains statistics or data \\
\textit{Controversial}    & The statement contains surprising or potentially controversial claims - e.g. a conspiracy theory \\
\textit{Opinion}& The statement contains claims about a person's subjective opinion or idea about something   \\
\textit{Private Life}& The statement contains claims about a person's private life - e.g. date of birth, relationship status.\\
\textit{Scientific}  & The statement contains technical or scientific claims    \\
\textit{Historical}  & The statement contains claims about general or historical facts that are not common knowledge    \\
\textit{Other}  & The statement requires a citation for reasons not listed above (please describe your reason in a sentence or two)    \\
\bottomrule
&\\
\multicolumn{2}{l}{\textbf{Reasons why citations are not needed}}   \\ \midrule
\textit{Common Knowledge} & The statement only contains common knowledge - e.g. established historical or observable facts   \\
\textit{Main Section}& The statement is in the lead section and its content is referenced elsewhere in the article \\
\textit{Plot}   & The statement is about a plot or character of a book/movie that is the main subject of the article    \\
\textit{Already Cited}    & The statement only contains claims that have been referenced elsewhere in the paragraph or article    \\
\textit{Other}  & The statement does not require a citation for reasons not listed above (please describe your reason in a sentence or two)\\
\bottomrule
\end{tabular}}
\caption{A taxonomy of Wikipedia verifiability: set of reasons for adding and not adding a citation. This taxonomy is the result of a qualitative analysis of various sources of information regarding Wikipedia editors' referencing behavior.}
\label{tab:citation_taxonomy}
\end{table*}

\vspace{2pt}
\noindent\textbf{Fact Checking and Verification.} 
Automated verification and fact-checking efforts are also relevant to our task of computationally understanding verifiability on Wikipedia.
Fact checking is the process of assessing the veracity of factual claims ~\cite{DBLP:journals/lre/SauriP09}. Long et al.~\cite{DBLP:conf/ijcnlp/LongLXLH17} propose TruthTeller 
computes annotation types for all verbs, nouns, and adjectives, which are later used to predict the truth of a clause or a predicate. Stanovsky et al.~\cite{DBLP:conf/acl/StanovskyEPDG17} build upon the output rules from TruthTeller and use those as features in a supervised model to predict the factuality label of a predicate. Chung and Kim~\cite{DBLP:journals/oir/ChungKK10} assess source credibility through a questionnaire and a set of measures (e.g. informativeness, diversity of opinions, etc.). The largest fact extraction and verification dataset FEVER~\cite{DBLP:conf/naacl/ThorneVCM18} constructs pairs of factual snippets and paragraphs from Wikipedia articles which serve as evidence for those factual snippets. However,these approaches cannot be applied in our case because they make the assumption that any provided statement is of factual nature.

Research on the automated fact detectors in political discourse \cite{konstantinovskiy2018towards,hassan2017toward,CLEF2018} is the work in this domain that is most closely related to ours. While these efforts have demonstrated the ability to effectively detect the presence of facts to be checked, they focus on the political discourse only, and they do not provide explanation for the models' prediction. In our work, we consider a wide variety of topics---any topic covered in Wikipedia---and design models able to not only detect claims, but also explain the reasons why those claims require citations.  

\section{A Taxonomy of Citation Reasons}\label{sec:taxonomy}

To train models for the \cnt{} and \crt{} tasks, we need to develop a systematic way to operationalize the notion of verifiability in the context of Wikipedia. 
There is currently no single, definitive taxonomy of reasons why a particular statement in Wikipedia should, or should not, have a supporting inline citation. We drew on several data sources to develop such a taxonomy using an inductive, mixed-methods approach. 



\vspace{2pt}
\noindent\textbf{Analyzing Citation Needed Templates.}
We first analyzed reasons Wikipedia editors provide when requesting an inline citation. Whenever an editor adds a \textit{citation needed} tag to a claim that they believe should be attributed to an external source, they have the option to specify a reason via a free-form text field. We extracted the text of this field from more than 200,000 \emph{citation needed} tags added by English Wikipedia editors and converted it into a numerical feature by averaging the vector representations of each sentence word, using Fasttext \cite{fasttext}. We then used k-means to cluster the resulting features into 10 clusters (choosing the number of clusters with the elbow method \cite{ketchen1996application}). Each cluster contains groups of consistent reasons why editors requested a citation. By analyzing these clusters
, we see that the usage of the ``reason'' field associated with the \textit{citation needed} tag does not consistently specify the reason why these tags are added. Instead, it is often used to provide other types of contextual information---for example, to flag broken links or unreliable sources, to specify the date when the tag was added, or to provide very general explanations for the edit. Therefore, we did not use this data to develop our taxonomy.

\vspace{2pt}
\noindent\textbf{Analyzing Wikipedia Citation Policies.}
As a next step, we analyzed documentation developed by the editor community to describe rules and norms to be followed when adding citations. We examined documentation pages in the English, French, and Italian language editions. Since each Wikipedia language edition has its own citation policies, we narrowed down the set of documents to analyze by identifying all subsidiary rules, style guides, and lists of best practices linked from the main Verifiability policy page, which exists across all three languages. Although these documents slightly differ across languages, they can be summarized into 28 distinct rules \footnote{The full guideline summary and the cluster analysis can be found here: \url{https://figshare.com/articles/Summaries_of_Policies_and_Rules_for_Adding_Citations_to_Wikipedia/7751027}}. Rules that we identified across these pages include a variety of types of claims that should usually or always be referenced to a source, such as claims of scientific facts, or any claim that is likely to be unexpected or counter-intuitive. These documentation pages also contain important guidance on circumstances under which it is appropriate to \emph{not} include an inline citation. For example, when the same claim is made in the lead section as well as in the main body of the article, it is standard practice to leave first instance of the claim unsourced.

\vspace{2pt}
\noindent\textbf{Asking Expert Wikipedians.}
To expand our \crtax, we asked a group of 36 Wikipedia editors from all three language communities (18 from English Wikipedia, 7 from French Wikipedia, and 11 from Italian Wikipedia) to annotate citations with reasons. Our experiment was as follows: we extracted sentences with and without citations from a set of Featured Articles and removed the citation metadata from each sentence. Using WikiLabels\footnote{\url{https://meta.wikimedia.org/wiki/Wiki_labels}}, an open-source tool designed to collect labeled data from Wikipedia contributors, we showed our annotators the original article with all citation markers removed and with a random selection of sentences highlighted. Editors were then asked to decide whether the sentence needed a citation or not (\cnt{} task), and to specify a reason for their choices (\crt{} task) in a free-text form. We clustered the resulting answers using the same methodology as above, and used these clusters to identify additional reasons for citing claims.

\vspace{2pt}
Our final set of 13 discrete reasons (8 for adding and 5 for not adding) is presented in Table \ref{tab:citation_taxonomy}. In Section \ref{sec:dataset}, we evaluate the accuracy of this taxonomy and use it to label a large number of sentences with citation-needed reasons. 

\section{Datasets}\label{sec:dataset}
In this Section, we show how we collected data to train models able to perform the \cnt{} task, for which we need sentences with binary citation/no-citation labels, and the \crt{} task, for which we need sentences labeled with one of the reason category from our taxonomy.

\subsection{\cnt{} Dataset}

Previous research~\cite{Forte:2018:IFO:3148330.3148347} suggests that the decision of whether or not to add a citation, or a \textit{citation needed} tag, to a claim in a Wikipedia article can be highly contextual, and that 
doing so reliably requires a background in editing Wikipedia and potentially domain knowledge as well. Therefore, to collect data for the \cnt{} task we resort to expert judgments by Wikipedia editors.

Wikipedia articles are rated and ranked into ordinal quality classes, from ``stub'' (very short articles) to ``Featured''
. Featured Articles\footnote{\url{https://en.wikipedia.org/wiki/Wikipedia:Featured_articles}} are those articles that are deemed as the highest quality by Wikipedia editors based on a multidimensional quality assessment scale \footnote{\url{https://en.wikipedia.org/wiki/Wikipedia:WikiProject_Wikipedia/Assessment\#Quality_scale}}. One of the criteria used in assessing Featured Articles is that the information in the article is \emph{well-researched}.\footnote{"[the article provides] a thorough and representative survey of the relevant literature; claims are verifiable against high-quality reliable sources and are supported by inline citations where appropriate." \url{https://en.wikipedia.org/wiki/Wikipedia:Featured_article_criteria}}
This criterion suggests that Featured Articles are more likely to consistently reflect best practices for when and why to add citations than lower-quality articles. The presence of \emph{citation needed} tags is an additional signal we can use, as it indicates that at least one editor believed that a sentence requires further verification.

We created three distinct datasets to train models predicting if a statement requires a citation or not\footnote{Unless otherwise specified, all data in the paper is from English Wikipedia}. Each dataset consists of: (i) \emph{positive instances} and (ii) \emph{negative instances}. Statements with an \emph{inline citation} are considered as \emph{positives}, and statements \emph{without an inline citation} and that appear in a \emph{paragraph with no citation} are considered as \emph{negatives}.

\vspace{2pt}
\noindent\textbf{Featured -- FA.} From the set of 5,260  \emph{Featured} Wikipedia articles we randomly sampled 10,000 positive instances, and equal number of negative instances.


\vspace{2pt}
\noindent\textbf{Low Quality (citation needed) -- LQN.}  In this dataset, we sample for statements from the 26,140 articles where at least one of the statements contains a \emph{citation needed} tag. 
The \emph{positive instances} consist solely of statements with \emph{citation needed} tags. 

\vspace{2pt}
\noindent\textbf{Random -- RND.} In the random dataset, we sample for a total of 20,0000 positive and negative instances from all Wikipedia articles. This provides an overview of how editors cite across articles of varying quality and topics. 


\begin{figure}
    \centering
        \includegraphics[width=\linewidth]{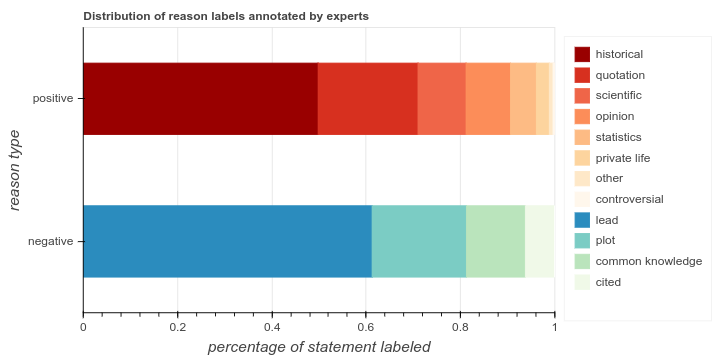}
        \caption{Distribution of labels assigned by Wikipedia Editors through the Wikilabels platform to characterize the reason why statements need citations.}\label{fig:wikilabels_results}
\end{figure}



\subsection{Citation Reason Dataset}
To train a model for the \crt{} task,
we designed a labeling task for Wikipedia editors in which they are asked to annotate Wikipedia sentences with both a binary judgment (citation needed/not needed) and the reason for that judgment using our \crtax. We used these annotations as ground truth for a larger-scale crowdsourcing experiment, where we 
asked micro-workers to select reasons for why \textit{positive} sentences require citations. 
We compared how well crowdworkers' judgments matched the Wikipedia editor judgments. Finally, we collected enough annotations to train machine learning algorithms.
 \begin{table*}[t]
\resizebox{0.85\linewidth}{!}{
\begin{tabular}{p{2cm}p{1.5cm}p{16cm}}
\textbf{Non-Expert judgment} & \textbf{Expert judgment}  & \textbf{Sentence extracted from Wikipedia Featured Article}            \\\toprule
\textit{historical}          & \textit{quotation}               & He argued that a small number of Frenchmen could successfully invade New Spain by allying themselves with some of the more than 15,000 Native Americans who were angry over Spanish enslavement   \\
\textit{life} & \textit{historical}                  & Actor Hugh Jackman is also a fan of the club, having been taken to Carrow Road as a child by his English mother, though he turned down an opportunity to become an investor in the club in 2010   \\
\textit{statistics}          & \textit{historical}                  & The act, authored by Ohio senator and former Treasury secretary John Sherman, forced the Treasury to increase the amount of silver purchased to 4,500,000 troy ounces (140,000 kg) each month     \\
\textit{quotation}    & \textit{historical}                  & "This stroke", said Clark, "will nearly put an end to the Indian War." Clark prepared for a Detroit campaign in 1779 and again in 1780, but each time called off the expedition because of insufficient men and supplies        \\
\bottomrule
\end{tabular}}
\caption{Example of sentences annotated with different categories by Wikipedia experts and Mechanical Turk contributors.}
\label{tab:mismatch_sentences}
\end{table*}

\subsubsection{Round 1: Collecting Data from Wikipedia Editors}

To collect ``expert'' annotations from Wikipedia editors on why sentences need citations, we proceeded as follows.

\vspace{2pt}
\noindent\textbf{Interface Design.}
We created a modified version of the free-text WikiLabels labeling task described in Section \ref{sec:taxonomy}. We selected random sentences from  Featured Articles, and removed citation markers when present.  
We presented the participants with the unsourced sentence highlighted in an article and asked them to label the sentence as needing an inline citation or not, and to specify a reason for their choice using a drop-down menu pre-filled with categories from our taxonomy. 
We recruited participants through mailing lists, social media and the English Wikipedia's Village pump (the general discussion forum of the English Wikipedia volunteer community).

\vspace{2pt}
\noindent\textbf{Results.}
We collected a total of 502 labels from 35 English Wikipedia editors.  Of the valid\footnote{Due to a bug in the system, not all responses were correctly recorded.} annotated sentences, 255 were labeled as needing a citation (\textit{positive}), and 80 as not needing a citation. Fig. \ref{fig:wikilabels_results} shows the breakdown of results by selected reason.

We found that the reason given for roughly 80\% of the \textit{positive} sentences 
is that they are "historical facts", "direct quotations", or "scientific facts". Furthermore, we observed that only a small percentage of participants selected the "Other" option, which suggests that our \crtax{} is robust and makes sense to editors, even when they are asked to provide these reasons outside of their familiar editing context.

 \subsubsection{Round 2: Collecting Data from non-Experts}\label{sec:crowdsourcing}
  We adapted the task in Round 1 to collect data from crowdworkers to train a \crt{} model.
 
\vspace{2pt}
\noindent\textbf{Task adaptation.}
 Adapting classification tasks that assume a degree of domain expertise to a crowdwork setting, where such expertise cannot be relied upon, can create challenges for both reliability and quality control. Crowdworkers and domain experts may disagree on classification tasks that require special knowledge~\cite{Sen:2015:TSA:2675133.2675285}. However, 
 Zhang et al.\cite{Zhang:2018:SRM:3184558.3188731} found that non-expert judgments about the characteristics of statements in news articles, such as whether a claim was well supported by the evidence provided, showed high inter-annotator agreement and 
 high correlation with expert judgments. In the context of our study, this suggests that crowdworkers may find it relatively easier to provide reasons for citations than to decide which sentences require them in the first place. 
 Therefore, we simplified the annotation task for crowdworkers
 to increase the likelihood of eliciting 
 high-quality judgments from non-experts. While Wikipedia editors were asked to both identify whether a sentence required citation and provide a reason, crowdworkers were only asked to provide a reason why a citation was needed.

\vspace{2pt}
\noindent\textbf{Experimental Setup.} 
We used Amazon Mechanical Turk for this annotation task. For each task, workers were shown one of 166 sentences that had been assigned citation reason categories by editors in round 1. Workers were informed that the sentence came from a Wikipedia article and that in the original article it contained a citation to an external source. Like editors in the first experiment, crowdworkers were instructed to select the most appropriate category from the eight citation reasons \ref{tab:citation_taxonomy}. 
Each sentence was classified by 3 workers, for a total of 498 judgments. For quality control purposes, only crowdworkers who had a history of reliable annotation behavior were allowed to perform the task. Average agreement between workers was 0.63\% (vs random 1/8 =0.125). 

\subsubsection{Comparing Expert and Non-Expert annotations}
\begin{figure}[t]
        \includegraphics[width=0.8\columnwidth]{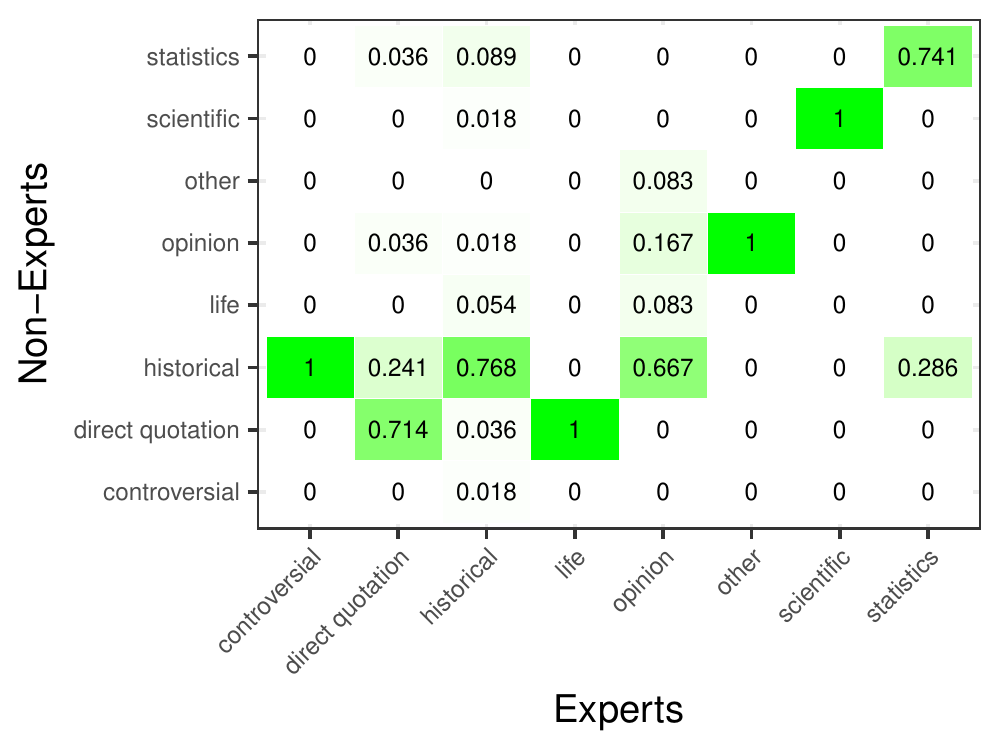}\vspace{-10pt}
\caption{ Confusion matrix indicating the agreement between Mechanical Turk workers ("non-experts") and Wikipedia editors ("experts"). The darker the square, the higher the percent agreement between the two groups}
\label{fig:comparison_charts}
\end{figure}

 The distribution of citation reasons provided by crowdworkers is shown in Fig. \ref{fig:cite_reason_dist_small}. The overall proportions are similar to that provided by Wikipedia editors in Round 1 (See Fig. \ref{fig:wikilabels_results}). Furthermore, the confusion matrix in Fig. \ref{fig:comparison_charts} indicates that crowdworkers and Wikipedia editors had high agreement on four of the five most prevalent reasons: \textit{historical, quotation, scientific} and \textit{statistics}. Among these five categories, non-experts and experts disagreed the most on \textit{opinion}. One potential reason for this disagreement is that identifying whether a statement is an opinion may require additional context (i.e. the contents of the preceding sentences, which crowdworkers were not shown). 

The confusion matrix in Fig. \ref{fig:comparison_charts}) shows the percentage of different kinds of disagreement---for example, that crowdworkers frequently disagreed with editors over the categorization of statements that contain \textit{"claims about general or historical facts."} To further investigate these results, we manually inspected a set of individual sentences with higher disagreement between the two groups. We found that in these cases the reason for the disagreement was due to a sentence containing multiple types of claims, e.g. a historical claim and a direct quote (see Table \ref{tab:mismatch_sentences}). This suggests that in many cases these disagreements were not due to lower quality judgments on the part of the crowdworkers, but instead due to ambiguities in the task instructions and labeling interface. 

 \subsubsection{The  \textit{Citation Reason Corpus}: Collecting Large-scale Data} 
\begin{figure}[t]
	\centering
	\includegraphics[width=\columnwidth]{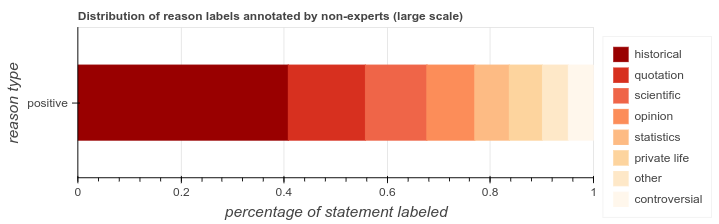}
	\caption{Citation reason distribution from the small-scale (166 sentences) crowdsourcing experiment.}
	\label{fig:cite_reason_dist_small}
\end{figure}

 Having verified the agreement between Wikipedia editors and crowdworkers, we can now reliably collect larger scale data to train a \crt{} model.  
 To this end, we sampled 4,000 sentences that contain citations from Featured articles, and asked crowdworkers to annotate them with the same setup described above (see Sec \ref{sec:crowdsourcing}). The distribution of the resulting judgments is similar to Fig. \ref{fig:cite_reason_dist_small}: as in Round 1, we found that the top categories are the \textit{scientific},\textit{quotation} and \textit{historical} reasons.\footnote{Our \textit{Citation Reason corpus} is publicly available here: \url{https://figshare.com/articles/Citation_Reason_Dataset/7756226}.}
\section{A \cnt{} Model}\label{sec:citation_needed_approach}

We design a classifier to detect when a statement needs a citation. We 
propose
a neural based Recurrent Neural Network (RNN) approach with varying representations of a statement, and compare it with a baseline feature-based model.  

\subsection{Neural Based \cnt{} Approach}

We propose a neural based model, which uses a recurrent neural network (RNN) with GRU cells\cite{cho2014learning} to encode statements for classification. We distinguish between two main modes of statement encoding: (i) vanilla RNN, fed with 2 different representations of a sentence (words and section information, indicated with $RNN^{w}$ and $RNN^{+S}$), and (ii) RNN with global attention $RNN_a$ (with similar representation).

\subsubsection{{Statement Representation}} For a given Wikipedia sentence, for which we want to determine its citation need, we consider the words in the statement and the section the statement occurs in. To feed the network with this information, we transform sentence words and section information into features, or representations. Through the word representation we aim at capturing cue words or phrases that are indicators of a statement requiring a citation. Section representation, on the other hand, allows us to encode information that will play a crucial role in determining the \crt{} later on.

\noindent\textbf{Word Representation.} We represent a statement as a sequence of words $s = (w_1, \ldots, w_n)$. We use GloVe pre-trained word embeddings~\cite{pennington2014glove} to represent the words in $s$. Unknown words are randomly initialized in the word embedding matrix $W_{glove}\in\mathbb{R}^{k\times 100}$, where $k$ is the number of words in the embedding matrix. 

\noindent\textbf{Section Representation.} The section in which the statement occurs in a Wikipedia article is highly important. The guidelines for \emph{inline citations} suggest that when a statement is in the \emph{lead section}, and that is referenced elsewhere in the article, editors should avoid multiple references \footnote{\url{https://en.wikipedia.org/wiki/Wikipedia:Manual_of_Style/Lead_section}}. Additionally, since  sections can be seen as a topically coherent group of information, the reasons for citation will vary across sections (e.g. \emph{``Early Life''}). We train the \emph{section embedding} matrix $W_{S}\in \mathbb{R}^{l\times 100}$, and use it in combination with $W_{glove}$, where $l$ is the number of sections in our dataset.

\subsubsection{{Statement Classification}}
We use 2 types of Recurrent Neural Networks to classify the sentence representations.

\vspace{2pt}
\noindent\textbf{Vanilla RNNs.} RNNs encode the individual words into a hidden state $h_t=f(w_t, h_{t-1})$, where $f$ represents GRU cells~\cite{cho2014learning}. The encoding of an input sequence from $s$ is dependent on the previous hidden state. This dependency based on $f$ determines how much information from the previous hidden state is passed onto $h_t$. For instance, in case of GRUs, $h_t$ is encoded as following:
\begin{equation}
	h_t = (1 - z_t) \odot h_{t-1} + z_t \odot \tilde{h}_t
\end{equation}
where, the function $z_t$ and $\tilde{h}_t$ are computed as following:
\begin{align}
	z_t = \sigma( W_z w_t + U_z h_{t-1} + b_z) \\
	\tilde{h}_t = \tanh\left(W_h w_t + r_t \odot (U_h h_{t-1} + b_h)\right) \\
	r_t = \sigma( W_r w_t + U_r h_{t-1} + b_r)
\end{align}
The RNN encoding allows us to capture the presence of words or phrases that incur the need of a citation. Additionally, words that do not contribute in improving the classification accuracy are captured through the model parameters in function $r_t$, allowing the model to ignore information coming from them. 

\vspace{2pt}
\noindent\textbf{RNN with Global Attention -- $RNN_a$.} As we will see later in the evaluation results, the disadvantage of vanilla RNNs is that when used for classification tasks, the classification is done solely based on the last hidden state $h_N$. For long statements this can be problematic as the hidden states, respectively the weights are highly compressed across all states and thus cannot capture the importance of the individual words in a statement. 

Attention mechanisms~\cite{bahdanau2014neural} on the other hand have proven to be successful in circumventing this problem. The main difference with standard training of RNN models is that all the hidden states are taken into account to derive a \emph{context vector}, where different states contribute with varying weights, or known with \emph{attention weights} in generating such a vector. 

\begin{figure}[t]
	\centering
	\includegraphics[width=.85\columnwidth,height=150px]{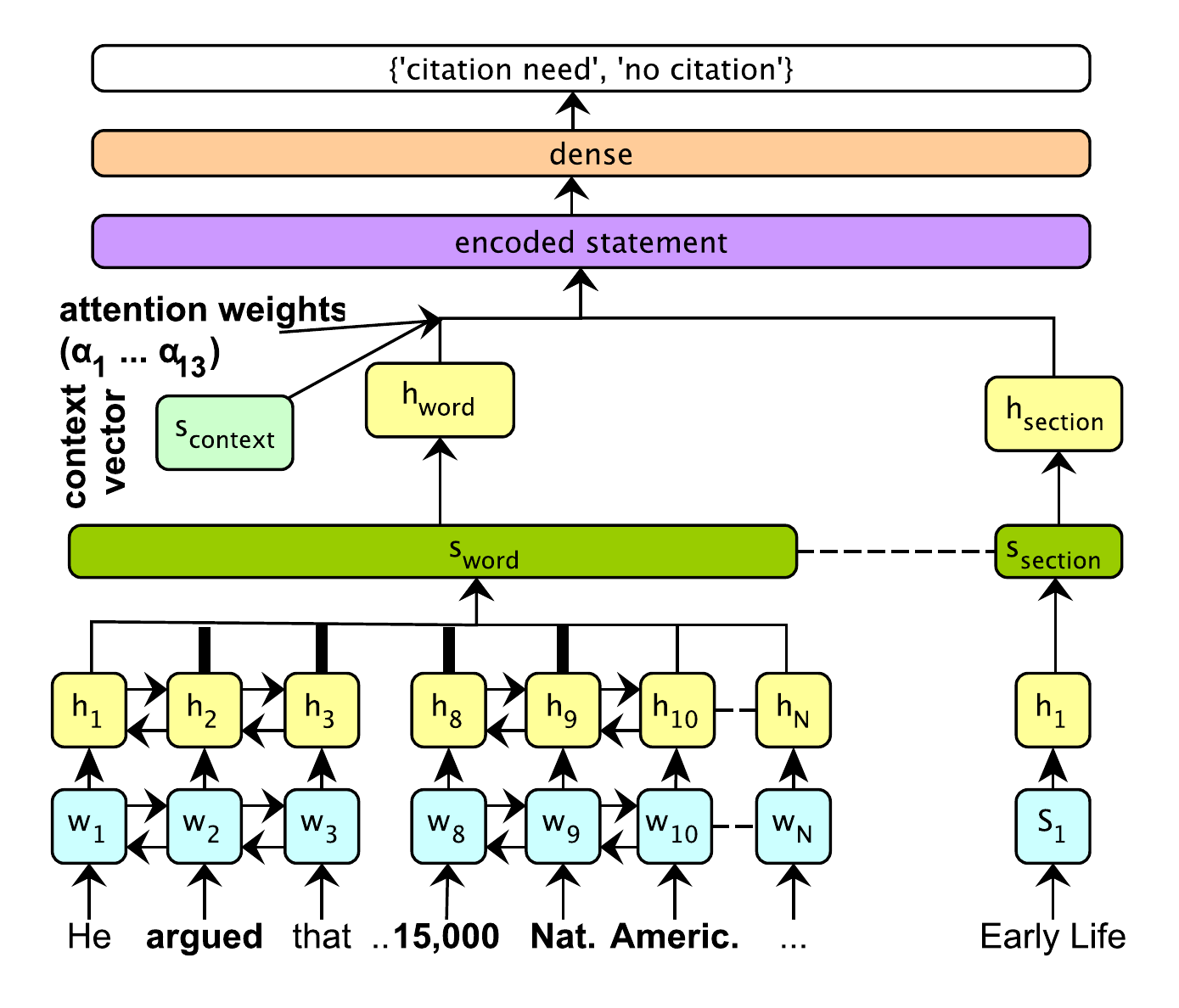}
	\caption{\cnt{} model with RNN and global attention, using both word and section representations.}
	\label{fig:citation_needed_model}
\end{figure}

Fig.~\ref{fig:citation_needed_model} shows the $RNN^{+S}_a$ model we use to classify a statement. We encode the statement through a bidirectional RNN based on its word representation, while concurrently a separate RNN encodes the section representation. Since not all words are equally important in determining if a statement requires a citation, we compute the \emph{attention weights}, which allow us to compute a \emph{weighted representation} of the statement based on the hidden states (as computed by the GRU cells) and the \emph{attention weights}. Finally, we \emph{concatenate} the weighted representation of the statement based on its words and section, and push it through a dense layer for classification.

The vanilla RNN, and the varying representations can easily be understood by referring to Fig.~\ref{fig:citation_needed_model}, by simply omitting either the section representation or the attention layer.

\subsubsection{Experimental Setup} 
We use Keras~\cite{chollet2015keras} with Tensorflow as backend for training our RNN models. We train for 10 epochs (since the loss value converges), and we set the  batch size to 100. We use Adam~\cite{kingma2014adam} for optimization, and optimize for \emph{accuracy}. We set the number of dimensions to 100 for hidden states $h$, which represent the words or the section information. 

We train the models with 50\% of the data and evaluate on the remaining portion of statements.

\begin{figure*}[t]
    \centering
    \includegraphics[width=0.8\textwidth]{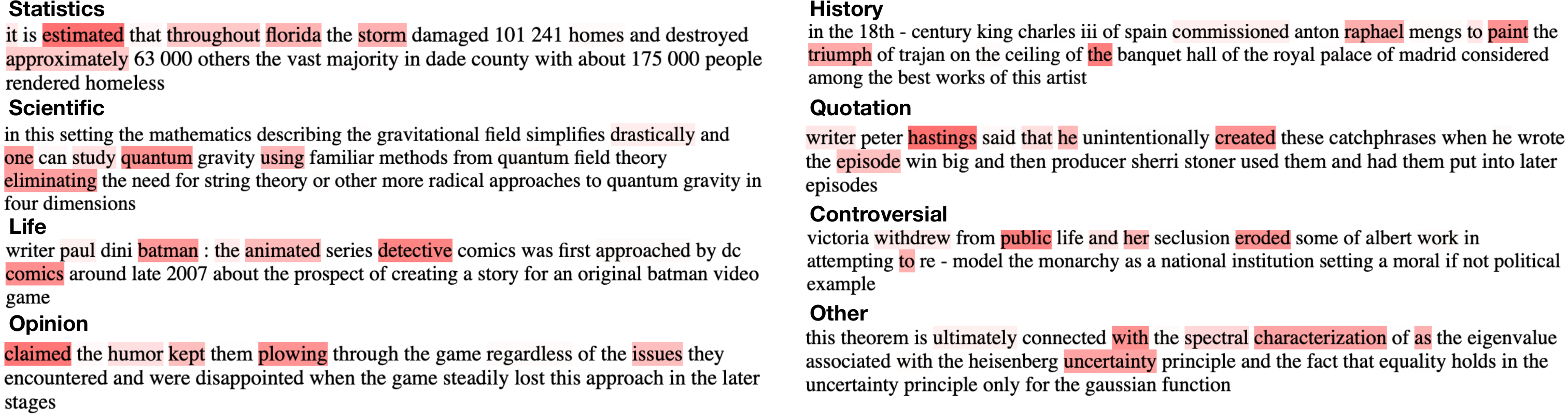}
    \caption{Attention mechanism for $RNN_a^{+S}$ visualizing the focus on specific words for the different citation reasons. It is evident that the model is able to capture patterns similar to those of human annotators (e.g. \emph{``claimed''} in the case of \emph{opinion}.)}
    \label{fig:citation_reason_attn_viz}
\end{figure*}
\begin{table}[]
\resizebox{\linewidth}{!}{
\begin{tabular}{ll|ll|ll}
\multicolumn{2}{l|}{\textbf{FA}} & \multicolumn{2}{l|}{\textbf{LQN}} & \multicolumn{2}{l}{\textbf{RND}}   \\ \hline
\textit{Section}& -0.621    & \textit{underline}   & 0.054  & \textit{say}    & 0.084  \\
\textit{say}   & 0.107& \textit{say}   & 0.0546    & \textit{underline}    & 0.0842  \\
\textit{underline}   & 0.107& \textit{believe}     & 0.042    & \textit{Section} & -0.072 \\
\textit{realize}     & 0.068     & \textit{disagree}    & 0.040    & \textit{report} & 0.062  \\
\textit{suggest}     & 0.068     & \textit{claim} & 0.039    & \textit{tell}   & 0.062 
\end{tabular}}\caption{Point-Biserial Correlation Coefficient between citation need labels and individual feature values}\label{tab:correlations}
\end{table}

\subsection{Feature-based Baselines}

As we show in Table~\ref{tab:citation_taxonomy}, where we extract the reasons why statements need a citation based on expert annotators, the most common reasons (e.g. \emph{statistics}, \emph{historical}) can be tracked in terms of specific \emph{language frames} and \emph{vocabulary use} (in the case of \emph{scientific} claims). Thus, we propose two baselines, which capture this intuition of language frames and vocabulary. From the proposed feature set, we train standard supervised models and show their performance in determining if a statement requires a citation.

    %

\subsubsection{Dictionary-Based Baseline -- Dict.} In the first baseline, we consider two main groups of features. First, we rely on a set of lexical dictionaries that aim in capturing words or phrases that indicate an activity, which when present in a statement would imply the necessity of a citation in such cases. We represent each statement as a feature vector where each element correspond to the frequency of a dictionary term in the statement.

\vspace{2pt}
\noindent\textbf{Factive Verbs.}  The presence of \emph{factive verbs}~\cite{kiparsky1968fact} in a statement presumes the truthfulness of information therein.

\vspace{2pt}
\noindent\textbf{Assertive Verbs.} In this case, assertive verbs~\cite{hooper1974assertive} operate in two dimensions. First, they indicate an assertion, and second, depending on the verb, the credibility or certainty of a proposition will vary (e.g. \emph{``suggest''} vs. \emph{``insist''}). Intuitively, \emph{opinions} in Wikipedia fall in this definition, and thus, the presence of such verbs will be an indicator of opinions needing a citation.

\vspace{2pt}
\noindent\textbf{Entailment Verbs.} As the name suggests, different verbs entail each other, e.g. \emph{``refrain''} vs. \emph{``hesitate''}~\cite{berant2012efficient,karttunen1971implicative}. They are particularly interesting as the context in which they are used may indicate cases of \emph{controversy}, where depending on the choice of verbs, the framing of a statement will vary significantly as shown above. In such cases, Wikipedia guidelines strongly suggest the use of citations.

\vspace{2pt}
\noindent\textbf{Stylistic Features.} Finally, we use the frequency of the different POS tags in a statement. POS tags have been successfully used to capture linguistic styles in different genres~\cite{DBLP:journals/coling/PetrenzW11}. For the different citation reasons (e.g. \emph{historical, scientific}), we expect to see a variation in the distribution of the POS tags. 

    %
    
    \subsubsection{Word Vector-Based Baseline -- WV} Word representations have shown great ability to capture word contextual information, and their use in text classification tasks has proven to be highly effective~\cite{DBLP:conf/eacl/GraveMJB17}. In this baseline, our intuition is that we represent each statement by averaging the individual word representations from a pre-trained word embeddings~\cite{pennington2014glove}. Through this baseline we aim at addressing the cases, where the \emph{vocabulary use} is a high indicator of statements needing a citation, e.g. \emph{scientific} statements.

    \subsubsection{Feature Classifier}
    We use a Random Forest Classifier \cite{breiman2001random} to learn \cnt{} models based on these features. To tune the parameters (depth and number of trees), similar to the main deep learning models, we split the data into train, test and validation (respectively 50\%,30\% and 20\% of the corpus). We perform cross-validation on the training and test set, and report accuracy results in terms of F1 on the validation set.
\begin{figure*}[t]
    \begin{subfigure}[b]{0.4\textwidth}
  \includegraphics[width=0.9\textwidth]{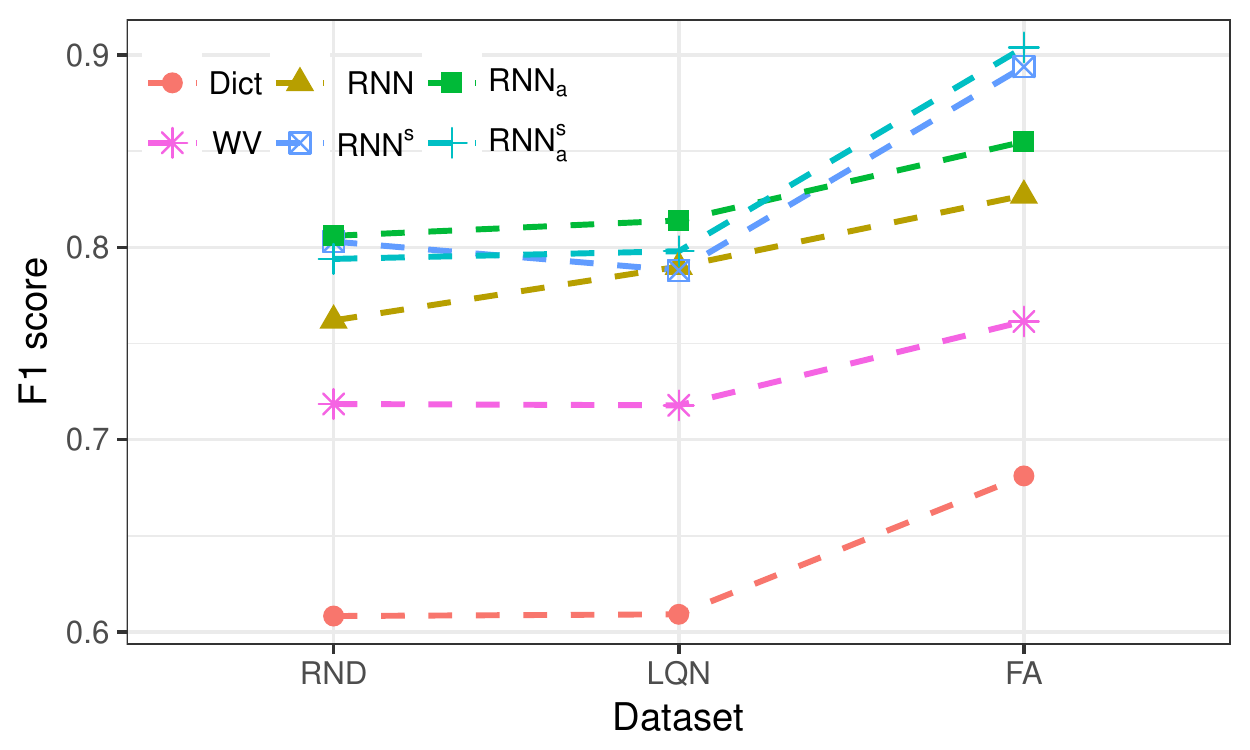}
  \label{fig:classification}
    \end{subfigure}
    \begin{subfigure}[b]{0.08\textwidth}
  \end{subfigure}
    ~ 
    \begin{subfigure}[b]{0.4\textwidth}
  \includegraphics[width=0.9\textwidth]{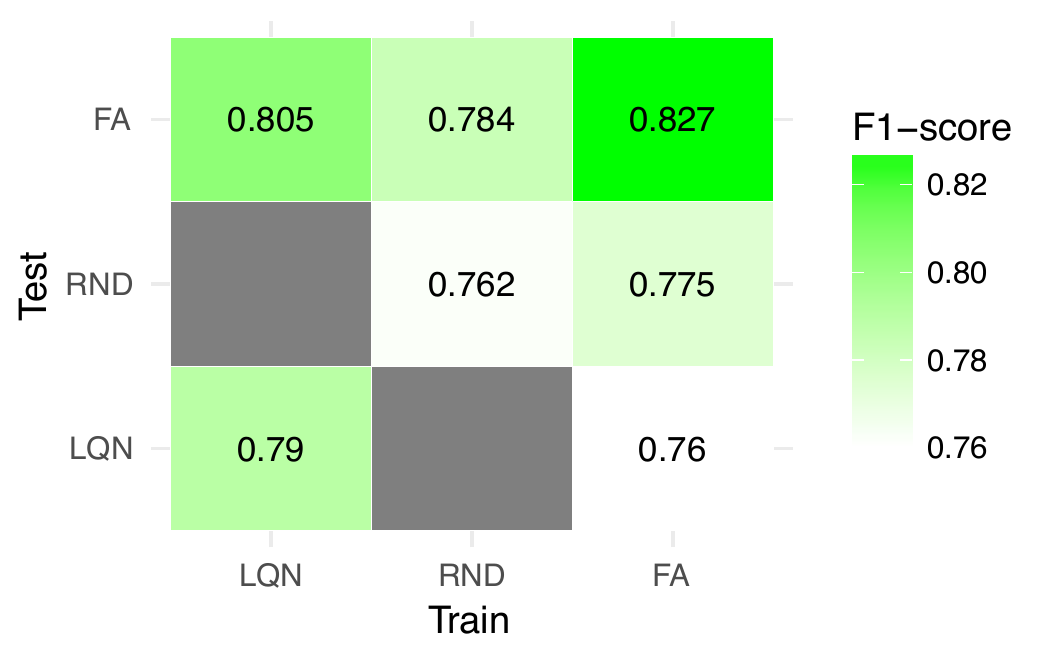}
  \label{fig:cross-classification}
    \end{subfigure}
\caption{(a) F1 score for the different \cnt{} detection models across the different dataset. (b) Confusion Matrix visualizing the accuracy (F1 score) of a \cnt{} model trained on Featured Articles and tested on other datasets, showing the generalizability of a model trained on Featured Articles only.}\label{fig:binary_plts}
\end{figure*}

\subsection{Citation Need Indicators}  
We analyze here how algorithms associate specific sentence features with the sentence's need for citations. 
    \subsubsection{Most Correlated Features}
    To understand which sentence features are more related to the need for citation, we compute the Point Biserial Correlation coefficient \cite{tate1954correlation} between the binary citation/no-citation labels and the frequency of each word in the baseline dictionary of each sentence, as well as the \textit{Section} feature.
    
    We report in Table \ref{tab:correlations} the top-5 most correlated features for each dataset. In featured articles, the most useful features to detect statements needing citation is the position of the sentence in the article, i.e. whether the sentence lies in the lead section of the article. This might be due to the fact that FA are the result of a rigorous formal process of iterative improvement and assessment according to established rubrics~\cite{Viegas2007}, and tend to follow the best practices to write the lead section, i.e. including general overview statements, and claims that are referenced and further verified in the article body. In the \textit{LQN} dataset we consider as ``positives" those sentences tagged as \textit{Citation Needed}. Depending on the article,  these tags can appear in the lead section too, thus explaining why the \textit{Section} feature is not discriminative at all for this group of sentences. Overall, we see that \emph{report verbs}, such as \textit{say, underline, claim} are high indicators of the sentence's need for citations.
\subsubsection{Results from Attention Mechanisms in Deep Learning}
Fig.~\ref{fig:citation_reason_attn_viz} shows a sample of positive statements from Featured Articles grouped by citation reason. The words are highlighted based on their \emph{attention weight} from the $RNN_a^{+S}$ model. The highlighted words show very promising directions. It is evident that the $RNN_a^{+S}$ model attends with high weights words that are highly intuitive even for human annotators. For instance, if we consider the \emph{opinion} citation reason, the highest weight is assigned to the word \emph{``claimed''}. This is case is particularly interesting as it capture the \emph{reporting verbs}~\cite{recasens2013linguistic} (e.g. \emph{``claim''}) which are common in opinions. In the other citation reasons, we note the \emph{statistics} reason, where similarly, here too, the most important words are again verbs that are often used in reporting numbers. For statements that are \emph{controversial}, the highest attention is assigned to words that are often used in a negative context, e.g. \emph{``erode''}. However, here it is interesting, that the word \emph{``erode''} is followed by context words such as 
\emph{``public''} and \emph{``withdrew''}. From the other cases, we see that the attention mechanism focuses on domain-specific words, e.g \emph{scientific} citation reason. 
\begin{table}[]
\resizebox{\linewidth}{!}{
\begin{tabular}{p{2.8cm}p{1.8cm}p{1.8cm}p{1.8cm}}
 & \textbf{no citation} & \textbf{citation} & \textbf{average} \\ \hline
\textit{individual editor} & 0.608  &\textbf{0.978}  & 0.766 \\
$RNN_a^{+S}$ & \textbf{0.902}  & 0.905  & \textbf{0.904}\\
\end{tabular}}
\caption{Accuracy (F1 score) of \cnt{} classification models on Featured Article vs individual expert editor annotations on the same set of Featured Articles.}
\vspace{-20pt}
\label{tab:human_machine}
\end{table}

\subsection{Evaluating the \cnt{} model}
In this section, we focus on assessing the performance of our model at performing the \cnt{} task, its generalizability, and how its output compares with the accuracy of human judgments.

    \subsubsection{Can an Algorithm Detect Statements in Need of a Citation?}
    We report the classification performance of models and baselines on different datasets in Fig. \ref{fig:binary_plts}. 
    
    Given that they are highly curated, sentences from Featured Articles are much easier to classify than sentences from random articles: the most accurate version of each model is indeed the one trained on the Featured Article dataset. 
    
    The proposed RNN models outperform the featured-based baselines by a large margin. We observe that adding attention information to a traditional RNN with GRU cells boosts performances by 3-5\%. As expected from the correlation results, the position of the sentence in an article, i.e. whether the sentence is in the lead section, helps classifying \cnt{} in Featured Articles only.
    \subsubsection{Does the Algorithm Generalize?}
    To test the generalizability of one the most accurate models, the RNN \cnt{} detection model trained on Featured Articles, we use it to classify statements from the \textit{LQN} and the \textit{RND} datasets, and compute the F1 score over such cross-dataset prediction. The cross-dataset prediction reaches a reasonable accuracy, in line with the performances models trained and tested on the other two noisier datasets. Furthermore, we test the performances of our $RNN_a$ model on 2 external datasets: the claim dataset from  Konstantinovskiy et al. \cite{konstantinovskiy2018towards}, and the CLEF2018 Check-Worthiness task dataset \cite{CLEF2018}. Both datasets are made of sentences extracted from political debates in UK and US TV-shows, labeled as positives if they contain facts that need to be verified by fact-checkers, or as negative otherwise. Wikipedia's literary form  is completely different from the political debate genre. Therefore, our model trained on Wikipedia sentences, cannot reliably detect claims in the fact-checking datasets above: most of the sentences from these datasets are outside our training data, and therefore the model tends to label all those  as negatives.
    \subsubsection{Can the Algorithm Match Individual Human Accuracy?}
    Our \cnt{} model performs better than individual Wikipedia editors under some conditions. Specifically, in our first round of expert citation labeling (Section \ref{sec:taxonomy} above), we observed that when presented with sentences from Featured Articles in the WikiLabels interface, editors were able to identify claims that already had a citation in Wikipedia with a high degree of accuracy (see Table \ref{tab:human_machine}), but they tended to \textit{over-label}, leading to a high false positive rate and lower accuracy overall compared to our model. There are several potential reasons for this. First, the editorial decision about whether to source a particular claim is, especially in the case of Featured Articles, an iterative, deliberate, and consensus-based process involving multiple editors. No single editor vets all the claims in the article, or decides which external sources to cite for those claims. Furthermore, the decisions to add citations are often discussed at length during the FA promotion process, and the editors involved in writing and maintaining featured articles often have subject matter expertise or abiding interest in the article topic, and knowledge of topic-specific citation norms and guidelines \cite{Forte:2014:DIS:2556288.2557072}. 
    By training on the entire corpus of Featured Articles, our model has the benefit of the aggregate of hundreds or thousands of editors' judgments of when (not) to cite across a range of topics, and therefore may be better than any individual editor at rapidly identifying general lexical cues associated with "common knowledge" and other statement characteristics that indicate citations are not necessary.
%
\small{
\begin{table}[t]
	\centering
	\begin{tabular}{l l l l l l l}
	\toprule
	& \multicolumn{3}{c}{\emph{pre-trained}} & \multicolumn{3}{c}{\emph{no pre-training}}\\
	\midrule
	& P & R & F1 	& P & R & F1\\
	\midrule
\emph{direct quotation} & \textbf{0.44} & \textbf{0.65} & \textbf{0.52} & 0.43 & 0.46 & 0.45\\
\emph{statistics} & 0.20 & \textbf{0.20} & \textbf{0.20} & \textbf{0.28} & 0.15 & 0.19\\
\emph{controversial} & \textbf{0.12} & \textbf{0.02} & \textbf{0.04} & 0.04 & 0.01 & 0.02\\
\emph{opinion} & \textbf{0.20} & 0.12 & 0.15 & 0.19 & 0.12 & 0.15\\
\emph{life} & 0.13 & 0.06 & 0.09 & \textbf{0.30} & 0.06 & \textbf{0.10}\\
\emph{scientific} & \textbf{0.62} & 0.56 & \textbf{0.59} & 0.54 & \textbf{0.58} & \textbf{0.56}\\
\emph{historical} & \textbf{0.56} & 0.67 & 0.61 & 0.54 & \textbf{0.74} & \textbf{0.62}\\
\emph{other} & 0.13 & 0.05 & 0.07 & \textbf{0.14} & \textbf{0.08} & \textbf{0.10}\\
\midrule
\emph{avg.} & 0.30 & \textbf{0.29} & \textbf{0.28} & \textbf{0.31} & 0.28 & 0.27\\
\bottomrule
	\end{tabular}	
	\caption{Citation reason prediction based on a pre-trained $RNN_a^{+S}$ model on the \emph{FA} dataset, and a $RNN_a^{+S}$ which we train only on the Citation Reason dataset.}
	\label{tab:citation_reason_pred}
\end{table}}
\begin{table}[t]
\resizebox{\columnwidth}{!}{
\begin{tabular}{lllllll}
\multicolumn{7}{c}{\textit{\textbf{Article Section}}}  \\ \hline
\textbf{quotation}   & \textbf{statistics}  & \textbf{controversial} & \textbf{opinion}  & \textbf{life}  & \textbf{scientific}  & \textbf{historical}   \\
reception  & history & history& reception& biography& description & history  \\
history& reception   & background & history  & history  & history & background   \\
legacy & legacy  & reception  & development & early life & taxonomy& abstract\\
production & abstract& legacy & production  & career  & habitat & aftermath      \\
biography  & description & aftermath  & background  & background & characteristics& life and career   \\
   & &  &  &   &  \\
\multicolumn{7}{c}{\textit{\textbf{Article Topics}}} \\ \hline
\textit{\textbf{quotation}} & \textit{\textbf{statistics}} & \textit{\textbf{controversial}} & \textit{\textbf{opinion}} & \textit{\textbf{life}} & \textit{\textbf{scientific}} & \textit{\textbf{historical}}    \\
videogame  & athlete & military conflict & videogame& athlete  & animal  & conflict   \\
athlete& settlement  & videogame  & athlete  & office holder  & fungus  & military person   \\
book   & videogame   & settlement & album& royalty  & plant   & royalty \\
officeholder  & infrastructure & athlete& single   & military & military unit  & office holder   \\
album  & country & royalty& book & artist& band& settlement    \\
\end{tabular}}
\caption{Most common article topics and article sections for the different citation reasons.}\label{table:types}
\end{table}

\section{A \crt{} Model}\label{sec:citation_reason}
In this Section, we analyze the \textit{Citation Reason Corpus} collected in Sec. \ref{sec:dataset}, and fine-tune the \cnt{} model to detect reasons why statements need citations.

\subsection{Distribution of Citation Reasons by Topic} 
Understanding if Wikipedia topics or article sections have different sourcing requirements may help contributors better focus their efforts.  
To start answering this question, we analyze citation reasons as a function of the article topic and the section in which the sentence occurs. We rely on DBpedia~\cite{auer2007dbpedia} to associate articles to topics and we show in Table \ref{table:types} the most topics and article sections associated with each citation reason. We note that the distribution of citation reasons is quite intuitive, both across types and sections. For instance, \emph{``direct quotation''} is most prominent in section \emph{Reception} (the leading section), which is intuitive, where the statements mostly reflect how certain \emph{``Athlete``}, \emph{``OfficeHolders``} have expressed themselves about a certain event.  Similarly, we see for \emph{``historical''} and \emph{``controversial''} the most prominent section is \emph{History}, whereas in terms of most prominent article types, we see that \emph{``MilitaryConflict''} types have the highest proportion of statements.

While the distribution of citation reasons is quite intuitive across types and sections, we find this as an important aspect that can be leveraged to perform targeted sampling of statements (from specific sections or types) which may fall into the respective citation reasons s.t we can have even distribution statements across these categories.

\subsection{Evaluating the \crt{} model} 
To perform the \crt{} task, we build upon the pre-trained model $RNN_a^{+S}$ in Fig.~\ref{fig:citation_needed_model}. We modify the $RNN_a^{+S}$ model by replacing the dense layer such that we can accommodate all the eight citation reason classes, and use a \emph{softmax}  function for classification.

The rationale behind the use of the pre-trained $RNN_a^{+S}$ model is that by using the much larger training statements from the binary datasets, we are able to adjust the model's weights to provide a better generalization for the more fine-grained citation reason classification. An additional advantage of using the model with the pre-trained weights is that in this way we can retain a large portion of the contextual information from the statement representation, that is, the context in which the words appear for statement requiring a citation. 

The last precaution we take in adjusting the $RNN_a^{+S}$ for \crt{} classification is that we ensure that the model learns a balanced representation for the different citation reason classes. 

Table~\ref{tab:citation_reason_pred} shows the accuracy of the pre-trained $RNN_a^{+S}$ model trained on 50\% of the Citation Reason dataset, and evaluate on the remaining statements. The pre-trained model has a better performance for nearly all citation reasons. It is important to note that due to the small number of statements in the Citation Reason dataset and additionally the number of classes, the prediction outcomes are not optimal. Our goal here is to show that the citation reason can be detected and we leave for future work a large scale evaluation.
\section{Discussion and Conclusions}\label{sec:discussion}
In this paper, we presented an end-to-end system to characterize, categorize, and algorithmically assess the verifiability of Wikipedia contents. In this Section we discuss the theoretical and practical implications of this work, as well as limitations and future directions.
\subsection{Theoretical Implications}
\textbf{A Standardization of Citation Reasons.} We used mixed methods to create and validate a \crtax. We then used this taxonomy to label around 4,000 sentences with reasons why they need to be referenced, and found that, in English Wikipedia, they are most often \textit{historical facts, statistics or data about a subject, or direct or reported quotations}.  Based on these annotations, we produced a Citation Reason corpus that we are making available to other researchers as open data\footnote{URL hidden for double blind submission}. While this taxonomy and corpus were produced in the context of a collaborative encyclopedia, given that they are not topic- or domain-specific, we believe they represent a resource and a methodological foundation for further research on online credibility assessments, in particular seminal efforts aiming to design controlled vocabularies for credibility indicators\cite{Zhang:2018:SRM:3184558.3188731}.

\vspace{2pt}
\noindent\textbf{Expert and Non-expert Agreement on Citation Reasons.} To create the verifiability corpus, we extended to crowdworkers a labeling task originally designed to elicit judgments from Wikipedia editors. We found that \textit{(non-expert) crowdworkers and (expert) editors agree about why sentences need citations in the majority of cases.} This result aligns with previous research~\cite{Kittur:2008:CUS:1357054.1357127}, demonstrating that while some kinds of curation work may require substantial expertise and access to contextual information (such as norms and policies), certain curation subtasks can be entrusted to non-experts, as long as appropriate guidance is provided. This has implications for the design of crowd-based annotation workflows for use in complex tasks where the number of available experts or fact-checkers doesn't scale, either because of the size of the corpus to be annotated or its growth rate.

\vspace{2pt}
\noindent\textbf{Algorithmic Solutions to the \cnt{} Task}. 
We used Recurrent Neural Networks to classify sentences in English Wikipedia as to whether they need a citation or not. We found that algorithms can effectively perform this task in English Wikipedia's Featured Articles, and generalize with good accuracy to articles that are not featured. We also found that, contrary to most NLP classification tasks, our \cnt{} model outperforms expert editors when they make judgments out of context. We speculate that this is because when editors are asked to make judgments as to what statements need citations in an unfamiliar article without the benefit of contextual information, and when using a specialized microtask interface that encourages quick decision-making, they may produces more conservative judgments and default to Wikipedia's general approach to verifiability---dictating that all information that's likely to be challenged should be verifiable, ideally by means of an inline citation. Our model, on the other hand, is trained on the complete Featured Article corpus, and therefore learns from the wisdom of the whole editor community how to identify sentences that need to be cited. 

\vspace{2pt}
\noindent\textbf{Algorithmic Solutions to the  \crt{} Task}
We made substantial efforts towards designing an interpretable \cnt{} model. In Figure~\ref{fig:citation_reason_attn_viz} we show that our model can capture words and phrases that describe citation reasons. 
To provide full explanations, we designed a model that can classify statements needing citations with a reason. To determine the citation reason, we modified the binary classification model $RNN_a^{+S}$ to predict the eight reasons in our taxonomy. We found that using the pre-trained model in the binary setting, we could re-adjust the model's weights to provide reasonable accuracy in predicting citation reasons. For citation reason classes with sufficient training data, we reached precision up to $P=0.62$. We also provided insights on how to further sample Wikipedia articles to obtain more useful data for this task.

\subsection{Limitations and Future Work}
Labeling sentences with reasons why they need a citation is a non-trivial task. Community guidelines for inline citations evolve over time, and are subject to continuous discussion: see for example the discussion about why in Wikipedia ``you need to cite that the sky is blue'' and at the same time ``you don't need to cite that the sky is blue'' \footnote{\tiny \url{https://en.wikipedia.org/wiki/Wikipedia:You_do_need_to_cite_that_the_sky_is_blue}}.
For simplicity, our \crt{} classifier treats citation reason classes as mutually exclusive. However, in our crowdsourcing experiment, we found that, for some sentences, citation reasons are indeed not mutually exclusive. In the future, we plan to add substantially more data to the verifiability corpus, and build multi-label classifiers as well as annotation interfaces that can account for fuzzy boundaries around citation reason classes. 

In Sec. \ref{sec:citation_needed_approach} we found that, while very effective on Wikipedia-specific data, our \cnt{} model is not able to generalize to fact-checking corpora. Given the difference in genre between the political discourse in these corpora, and the Wikipedia corpus, this limitation is to be expected.  We explored, however, two other generalizability dimensions: domain expertise and language. We demonstrated that, for this task, annotation can be effectively performed by non-experts, facilitating the solution of this task at scale and distributing it beyond expert communities. Moreover, we built a general multilingual taxonomy by evaluating policies from different Wikipedia language communities, and by testing its effectiveness with expert contributors from English, Italian, and French Wikipedia.

More broadly, this work is designed for multilingual generalizability. In the future, we aim to replicate the large annotation efforts across languages. This should be fairly straight-forward, since Featured Articles exist in 163 Wikipedia language editions\footnote{\url{https://www.wikidata.org/wiki/Q16465}}. 
Moreover, the RNN model can be fed with word vectors such as fasttext\cite{fasttext}, which now exist in more than 80 languages \cite{bojanowski2016enriching} and that one can re-train with any language from a Wikipedia project. 

Finally, in this study we consider the application of verifiability policies to a static snapshot of Wikipedia articles, not taking into account their revision history. We also used general text features, and limited the encyclopedic-specific features to the \textit{main section} feature. We expect that the distribution of citation reasons may vary over the course of an article's development, as a function of how controversial or disputed particular sections are. Performing this analysis is beyond the scope of the current study, but it might surface important exceptions to our findings or uncover interesting editing dynamics. 


\subsection{Practical Implications} 
Our study also has practical implications for the design of collaborative systems to support information verifiability and fact-checking. Our results of a robust agreement between non-experts and experts around citation reasons suggests that this type of task can be effectively crowdsourced, potentially allowing systems to recruit non-experts to triage or filter unverified statements based on model predictions, and allowing subject-matter experts to focus their attention on identifying reliable sources for these statements, or improving articles or topics with the highest rate of unreferenced factual claims. On Wikipedia, existing microtasking tools  such as CitationHunt\footnote{https://tools.wmflabs.org/citationhunt/} could be modified to surface unverified claims that have not yet been flagged \textit{citation needed} by humans and provide users with model predictions to guide their research, while also allowing users to provide feedback on those predictions to refine and improve the model. The model could also be used to surface article or statement-level information quality indicators to Wikipedia readers, using scores or lightweight visualizations, as suggested by Forte and others~\cite{Forte:2014:DIS:2556288.2557072,Adler:2008:ATW:1822258.1822293}, to support digital literacy and to allow readers to make more informed credibility assessments. Finally, downstream re-users of Wikipedia content, such as search engines and social media platforms, could also draw on model outputs to assign trust values to the information they extract or link to.

Beyond Wikipedia, our work complements existing attempts to assess how experts and non-experts assess the credibility of digital information,~\cite{Zhang:2018:SRM:3184558.3188731} and suggests that it is possible to develop robust \textit{verifiability taxonomies} and automated systems for identifying unverified claims in complex information spaces even without substantial domain knowledge. Such capabilities could support large-scale, distributed fact checking of online 
content, making the internet more robust against the spread of misinformation and increasing the overall information literacy of internet users.

\section*{Acknowledgment:} We would like to thank the community members of the English, French and Italian Wikipedia for helping with data labeling and for their precious suggestions, and Bahodir Mansurov and Aaron Halfaker from the Wikimedia Foundation, for their help  building the WikiLabels task. This work is partly funded by the ERC Advanced Grant ALEXANDRIA (grant no. 339233), and BMBF Simple-ML project (grant no. 01IS18054A). 
\bibliographystyle{ACM-Reference-Format}
\balance
\bibliography{bibliography}

\end{document}